\pdfoutput=1

\documentclass[twocolumn,floatfix,notitlepage,groupaddress,showpacs,nofootinbib]{revtex4-2}
\usepackage{times,color,xcolor,amsfonts,amssymb,amsbsy,amsthm,amsmath,graphics,graphicx,bm,bbm,makecell,mathtools,dsfont,upgreek,fancyhdr,multirow,newtxmath,subfigure}
\usepackage{soul}
\usepackage{natbib}

\DeclarePairedDelimiter{\ceil}{\lceil}{\rceil} 
\definecolor{pblue}{rgb}{0.11, 0.22, 0.73}

\begin{document}

\title{Composable end-to-end security of Gaussian quantum networks with untrusted relays}
% Running Head: END-TO-END SECURITY OF GAUSSIAN QUANTUM NETWORKS
\author{Masoud Ghalaii}
\thanks{These authors have contributed equally to this work.}
\author{Panagiotis Papanastasiou}
\thanks{These authors have contributed equally to this work.}
\author{Stefano Pirandola}
\affiliation{Department of Computer Science, University of York, York YO10 5GH, United Kingdom}

\maketitle
\thispagestyle{empty}

\textbf{
Gaussian networks are fundamental objects in network information theory. Here many senders and receivers are connected by physically motivated Gaussian channels while auxiliary Gaussian components, such as Gaussian relays, are entailed. Whilst the theoretical backbone of classical Gaussian networks is well established, the quantum analogue is yet immature. Here, we theoretically tackle composable security of arbitrary Gaussian quantum networks (quantum networks), with generally untrusted nodes, in the finite-size regime. We put forward a general methodology for parameter estimation, which is only based on the data shared by the remote end-users. Taking a chain of identical quantum links as an example, we further demonstrate our study. Additionally, we find that the key rate of a quantum amplifier-assisted chain can ideally beat the fundamental repeaterless limit with practical block sizes. However, this objective is practically questioned leading the way to new network/chain designs. 
}

\bigskip
{\bf\large{Introduction}}

$-\log_2(1-\eta)$, where $\eta$ is the channel's transmissivity, is the maximum fundamental rate, in bits, at which two distant parties can distribute quantum bits, entanglement bits, or secret bits. This is known as the Pirandola-Laurenza-Ottaviani-Banchi (PLOB) bound and holds for any point-to-point protocol of quantum communication~\cite{Pirandola:PLOB2017}. 
Since the discovery of PLOB, vast efforts have been made to break its hindrance, e.g., by using quantum repeater chains~\cite{Pirandola:AQCrypt,Cao:QKDReview2022}. 
In fact to outclass the PLOB bound, it is necessary to insert in-the-middle quantum stations, which can also be set out in a non-chain configuration to build a quantum network withal. Thus, one ultimate goal is not only to surpass the PLOB bound, but also to branch out a network of quantum links that would enable simultaneous secure communication or key distribution between more than just a few pairs of users~\cite{Pirandola:EndtoEnd2019}. Such telecommunication networks can further develop to provide us with a future quantum internet for quantum-secure communications~\cite{Pirandola:AQCrypt,Cao:QKDReview2022} and distributing quantum computing~\cite{Kimble:QuInternet,Pirandola:QInternet,Razavi:Book2018,Pirandola:Review15}. 

\begin{figure}[t]
\vspace{+.1cm}
\includegraphics[scale=.4]{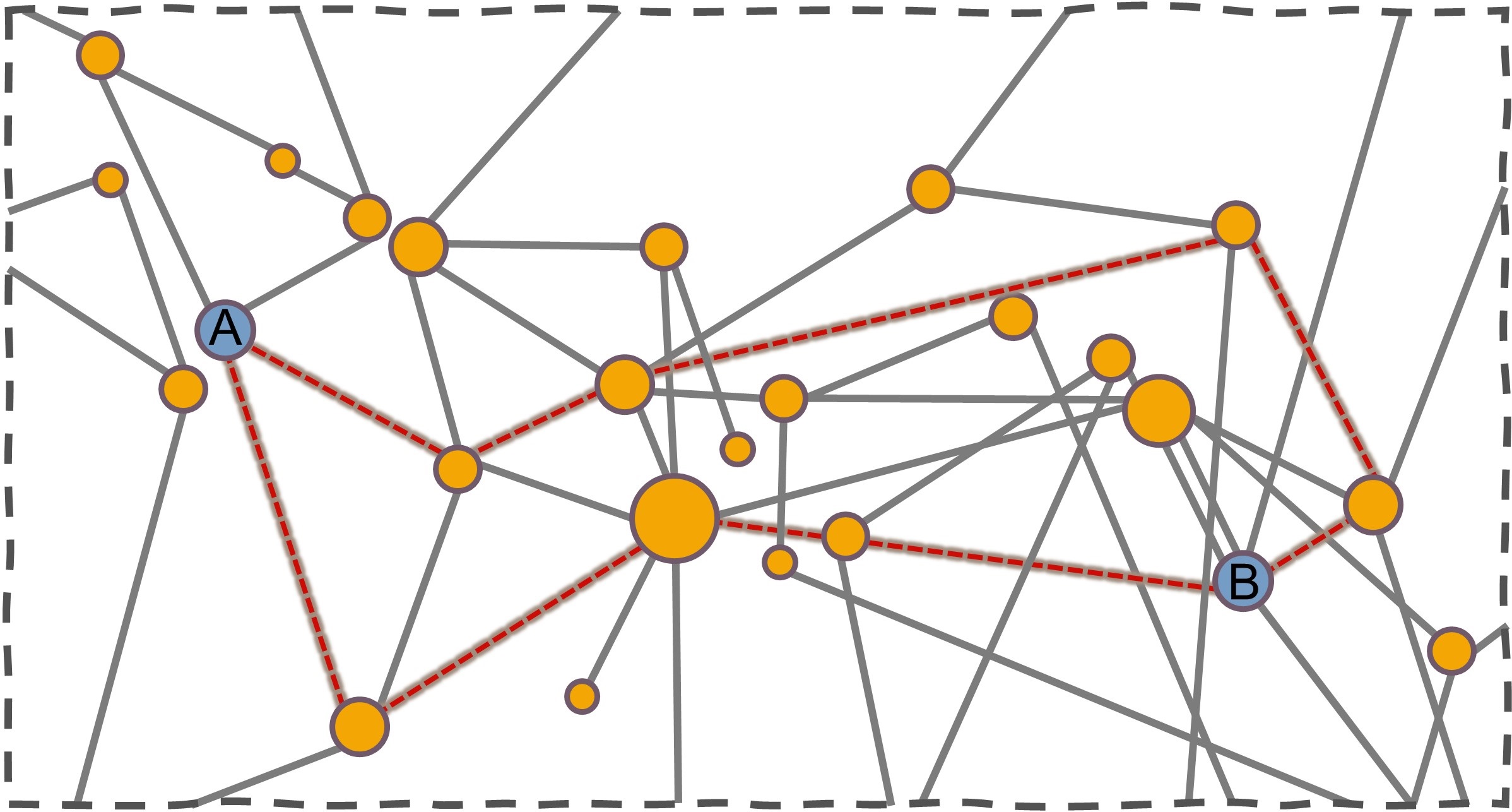}
\caption{{\bf Quantum communication network.} Two arbitrary end-users, Alice ($A$) and Bob ($B$), can communicate through diverse, not necessarily direct, routes that extend across intermediate untrusted sender-receiver pairs that act as relays (yellow nodes). Two possible routes are highlighted in red. The quantum network is Gaussian if the operations at the nodes and the channels associated with links are all Gaussian, so that the final state shared by Alice and Bob is Gaussian. More weakly, we also include the possibility of non-Gaussian post-selection operations which however project into a Gaussian state when they are successful (see text for more details). }
\label{fig:qnet}
\end{figure}

Gaussian networks, {\em inter alia}, are at the core of classical information theory, upon which concepts of communication networks have been developed~\cite{Cover}. Such networks, e.g., a large network of optical fibre links, have been studied and evolved in response to our continuous demand for data communications. They, as the name suggests, enjoy Gaussian signal assumptions and Gaussian links, where random variables with Gaussian probability density functions describe the channel noise. In addition, any other component, e.g., repeater relays, that makes  the process of data communications possible or alleviates it is Gaussian, such that none of the shared variables/distributions between users of the network becomes non-Gaussian. 
In this work, we put our focus on Gaussian quantum networks that benefit from Gaussian input signals, Gaussian quantum channels, and auxiliary Gaussian quantum devices. In particular, we study end-to-end security between two arbitrary users of a Gaussian quantum network who are generally linked via untrusted nodes (see Fig.~\ref{fig:qnet}). More weakly, as we explain later, we also admit some post-selection operations that are conditionally Gaussian, i.e., projecting into a Gaussian state when they are successful (discarding their output otherwise).

While examining a quantum network, not only is it fundamental to compute the relevant communication rates between arbitrary users, namely upper, lower and achievable rates, but it is crucial to evaluate composable key rates with a finite number of uses of the network. Evaluating the rate is possible by analysing the data statistics that the parties would obtain through the so-called parameter estimation (PE)~\cite{Tomamichel:2012,Furrer:FSize2012,Leverrier:FSize2017}. For a typical single link, PE analysis, which commonly refers to estimating {\em channel} parameters (loss and noise), is relatively straightforward~\cite{Papanastasiou:ComposMDI2017,Pirandola:CompCVQKD2021}. 
However, PE can become very challenging in large-scale quantum networks. For these reasons, we do not consider estimating channel parameters; instead, we use PE in a more general sense by directly estimating measurable quantities, e.g., the covariance matrix of the end parties. 

In the context of continuous-variable (CV) quantum key distribution (QKD), we show that any two users of a Gaussian quantum network can successfully extract composable secret keys from their local shared data, together with any classical public data that they might receive from other stations of the network. 
As mentioned above, an important point to remark is that we allow the network to deviate from being Gaussian, including the possibility to be conditionally Gaussian, i.e., described by a Gaussian state only after the success of a non-Gaussian, post-selection mechanism (feature which is needed for effective entanglement distillation~\cite{Eisert:NGEntDist2002,Eisert:NGEntDist2004,Ourjoumtsev:NGEntDist2007,Takahashi:NGEntDist2010,Kurochkin:NGEntDist2014}). In particular, this happens when non-deterministic quantum amplifiers are in use, where they sporadically fail to amplify~\cite{Ralph:QSNLA,Xiang:NLA,McMahon:Optimal_QS,Chrzanowski:MBNLA2014,Pandey:QAmps2013,Caves:PRA2012}. 
{We further investigate the use of such amplifiers in a linear quantum chain.}

\bigskip
{\bf\large{Results and Discussion}}

\begin{figure}
\vspace{+.1cm}
\includegraphics[scale=.45]{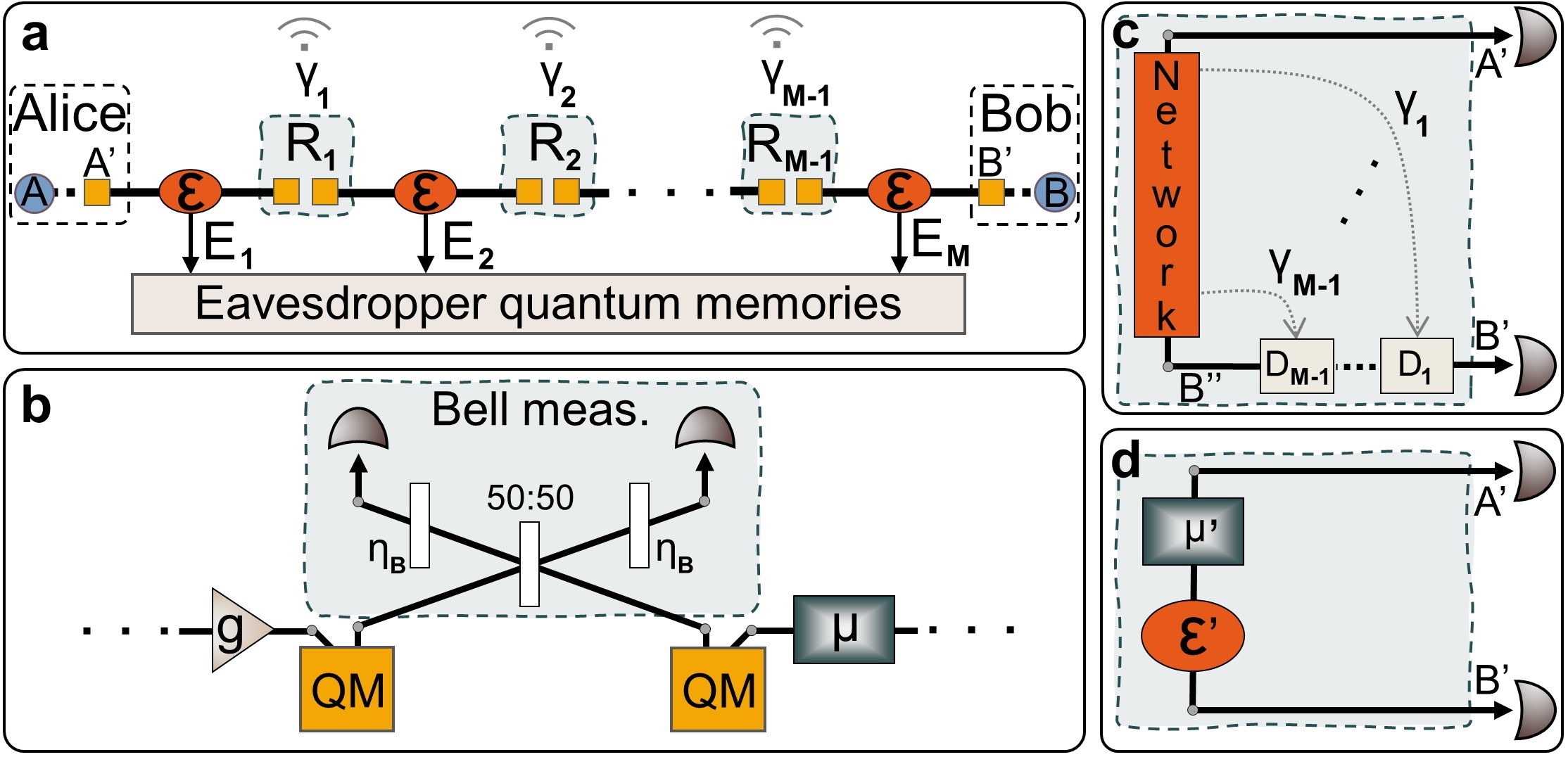}
\caption{{\bf A quantum communication chain within a network.} 
{\bf a,} An arbitrary route between Alice and Bob can be seen as a linear chain between them that consists of $M$ links and $M-1$ stations ($R_i$'s). {\bf b,} Each station can encompass a noiseless linear amplifier (NLA, $g$), a bipartite entanglement TMSV source ($\mu$), a couple of QMs, and a non-ideal {Bell detection}, whose loss is simulated by a couple of beam splitters with transmissivity $\eta_{\rm B}$. 
{\bf c,} A Gaussian quantum network provides the end parties with a Gaussian bipartite state, called the network state $\rho_{A'B'}$. Displacement operations need to be applied according to the information received form the stations, as shown in {\bf a}. 
{\bf d,} The effect of the Gaussian quantum network in {\bf c} can be simulated via a one-way Gaussian protocol with an equivalent source $\mu'$, and an equivalent channel, $\mathcal{E}'$, with loss $\eta'$ and excess noise $\xi'$.}
\label{fig:qchain}
\end{figure}

\smallskip
{\bf Gaussian quantum networks.} 
We consider the scenario where Alice and Bob are two arbitrary users of a quantum network, as sketched in Fig.~\ref{fig:qnet}; their objective is to remotely share a secret key. Let us assume that there are $M-1$ stations that relay signals from Alice to Bob through a specific route that is made of $M$ basic links. As Fig.~\ref{fig:qchain}a schematically illustrates, an arbitrary route can be seen as a chain of quantum links. 
It is also assumed that a powerful eavesdropper (Eve) may operate the intermediate stations and also store all the lost portion of the signals into her quantum memories (QMs). The relay stations may consist of several  components. For instance, they can be equipped with entanglement sources, such as two-mode squeezed vacuum (TMSV), quantum amplifiers, quantum memories, and a classical communication system; see Fig.~\ref{fig:qchain}b. Nonetheless, the key role of the relays is to connect adjacent links via joint Bell measurements, whose outcomes $\gamma_i$ (for $i=1,\dots,M-1$) are aired to Alice and Bob for local data processing. Note that, in case the relays operate differently from expected, this would reflect in high amount of noise in Alice and Bob's shared data. 

Figure~\ref{fig:qchain}c captures the role of the network in terms of quantum teleportation-stretching formalism~\cite{Pirandola:PLOB2017}. The network provides end-parties, Alice and Bob, with a bipartite (entangled) Gaussian state, which we call the {\em network state} $\rho_{A'B''|\{\gamma_i\}}$, {\em before} $\gamma_i$'s corrections are applied. 
We conventionally assume that the initial single links are of zero mean. However, execution of a relay, e.g., $R_1$, displaces the mean value of the state by an amount $f(\gamma_1)$ proportional to $\gamma_1$. In order to `correct' this a displacement operation, e.g., $\widehat{D}_1$, should be applied accordingly. Similar displacement operations are applied due to other relay outcomes that can all be postponed to one end. Thus, in this way, the mean value of the network state after $\gamma_i$ corrections, now described by $\rho_{A'B'}$, becomes independent of the $\gamma_i$'s (in fact we balance it to zero). 
Further, since displacements are local operations, the network state will have a covariance matrix (CM) $\mathbf{V}_{A'B'}=\mathbf{V}_{A'B''|\{\gamma_i\}}$, which is described in the normal form 
\begin{align}
\label{eq:CMNetState}
\mathbf{V}_{A'B'}= 
\left(\begin{array}{cc}
\mathsf{a} \mathbf{I} & \mathsf{c} \mathbf{Z} \\
\mathsf{c} \mathbf{Z} & \mathsf{b} \mathbf{I}
\end{array}\right),
~\left\{
\begin{array}
[c]{l}%
\mathbf{I}:=\mathrm{diag}(1,1),\\
\mathbf{Z}:=\mathrm{diag}(1,-1). 
\end{array}
\right.%
\end{align} 
Therefore, the network state supplies Alice and Bob with an overall two-mode Gaussian state that can be used to implement different one-way-like communication protocols. 

We remark that the Gaussianity assumption can be weakened because of conditioning, or post-selection, where relays can actually impose non-Gaussianity on the entire network, yet the system can be assumed {\em conditionally Gaussian}. This situation occurs because measurable quantities, such as CM elements, depend on relay measurement outcomes, which may vary for different sets of $\{\gamma_i\}$. This is similar to post-selection~\cite{Fiurasek:VirNLA2012,Li:PostSel2016} or discrete-alphabet protocols~\cite{Wilkinson:PostSel2020,Papanastasiou:DiscAlph2021} where, for example, the outcome set $\#1$ gives $\mathbf{V}_{A'B'}^{\#1}$, while the outcome set $\#2$ gives $\mathbf{V}_{A'B'}^{\#2}$ that differs from the CM associated to that of set $\#1$. Thus, one needs to build an average rate over all possible outcome sets. Therefore, the average state/CM would be non-Gaussian. Nevertheless, if in such situations we choose to discard the unsuccessful events, then the post-selected state between Alice and Bob is Gaussian.

\smallskip
{\bf Security reduction.} 
It is conceivable that the types of attack that eavesdroppers may apply on a multi-link quantum network can be more complex than the way they would attack conventional one-link protocols. 
For instance, in a quantum network a subset of the links that form the route from Alice to Bob may have correlations. In fact, Eve may adapt her attack on a link based on the information she has gained while attacking other, previous links. This generally defines a collective network attack, which has memory between the links but is memory-less between different uses of the network.
Such inter-link correlations are taken into account in the network state or, alternatively, its corresponding CM. This is due to the fact that we consider only the end-to-end Gaussian CM for the analysis.\footnote{{Note that we assume that the route is fixed. In the case it changes use-by-use, a more general attack would involve correlations between all the links that are overall used over multiple uses of the network.}}
As a requirement for our analysis, it is important to note that the CM of the network state is in normal form of Eq.~\eqref{eq:CMNetState}.

Nevertheless, there may also be correlations between subsequent uses of a route, which defines an even more powerful and general, coherent attack. Hence, we need to prove the security when the eavesdroppers develop inter-use correlations, i.e., when they apply a coherent attack.  
Our solution is to tackle this problem by reducing the Gaussian quantum network security to that of one-way protocols, for which optimality of collective Gaussian attacks has been proven ~\cite{GarciaPatron:Optim2006}. In this way, we reduce the complexity of the problem and extend the security analysis under collective attacks to coherent attacks. 

Assume that the end-nodes of the network, $A'$ and $B'$, remain in the Gaussian regime. We can seek for equivalent parameters of a single Gaussian channel that does the same job. In fact, the overall function of a Gaussian quantum network can be reduced to, and modelled by, a one-way Gaussian channel, with loss $\eta'$ and excess noise $\xi'$, applied to an equivalent source with modulation $\mu'$. See Fig.~\ref{fig:qchain}d, where we have that such equivalent parameters builds up a CM in normal form 
\begin{align}
\label{eq:CMEqivSt}
\mathbf{V}_{A'B'}^{\rm eqv}=
\left(\begin{array}{cc}
\mu' \mathbf{I} & \sqrt{\eta'(\mu'^2-1)} \mathbf{Z} \\
\sqrt{\eta'(\mu'^2-1)} \mathbf{Z} &  \eta'\mu'+1-\eta'+\xi' \mathbf{I}
\end{array}\right). 
\end{align}
It is straightforward to find the elements of the CM of the equivalent state, given by Eq.~\eqref{eq:CMEqivSt}, in the terms of the triplet $(\mathsf{a},\mathsf{b},\mathsf{c})$ in Eq.~\eqref{eq:CMNetState} that describes the network state $\rho_{A'B'}$; one can obtain 
\begin{align}
\begin{cases}
\mu'=\mathsf{a}, \\
\eta'=\mathsf{c}^2(\mathsf{a}^2-1)^{-1}, \\
\xi'=(\mathsf{a}+1)(\mathsf{b}-1)-\mathsf{c^2}. 
\end{cases}
\end{align}
Note that $\mathbf{V}_{A'B'}^{\rm eqv}$ is {\em bona fide}, i.e., $\mu'\geq 1$, $\eta'\leq 1$, and $\xi'\geq 0$, when the CM in Eq.~\eqref{eq:CMNetState} is {\em bona fide}, i.e., $\mathsf{a},\mathsf{b}\geq 1$ and $\mathsf{c}\leq \min \big\{\sqrt{\mathsf{a}^2-1}, \sqrt{(\mathsf{a}+1)(\mathsf{b}-1)} \big\}$~\cite{Pirandola:BonaFide2009}.

This means that the original collective network attacks can be extended to coherent network attacks where correlations could happen between different uses of the network. Consequently, the optimality of Gaussian attacks in typical one-way Gaussian protocols is extended to Gaussian quantum networks. It is therefore a reasonable assumption to consider Gaussian eavesdropping which is the optimal strategy in the presence of protocols based on Gaussian resources. For this reason, for our security analysis and composable study we consider network attacks that are collective and Gaussian.

%As a matter of fact, we assume a type of partial adaptivity within the network, but not temporal in terms of uses of the network, such that at each run the extra system of Eve interacts with that of users in a `fresh' way (reset memory). We take this as the meaning of collective Gaussian attacks in a quantum network setting. We shall present composable security of protocols against such attacks. 

%Furthermore, we note that optimality of collective Gaussian attacks has been proven for one-way protocols~\cite{GarciaPatron:Optim2006}. Whether such attacks are optimal in quantum networks remains open. However, we conjecture that a chain of links in a quantum network can be reduced to an equivalent one-way channel (through channel dilation and purification). If so, this would automatically prove the conjecture as well as the security against coherent attacks. ****TO change *** 

\begin{figure*}
\vspace{+.1cm}
\includegraphics[width=15cm]{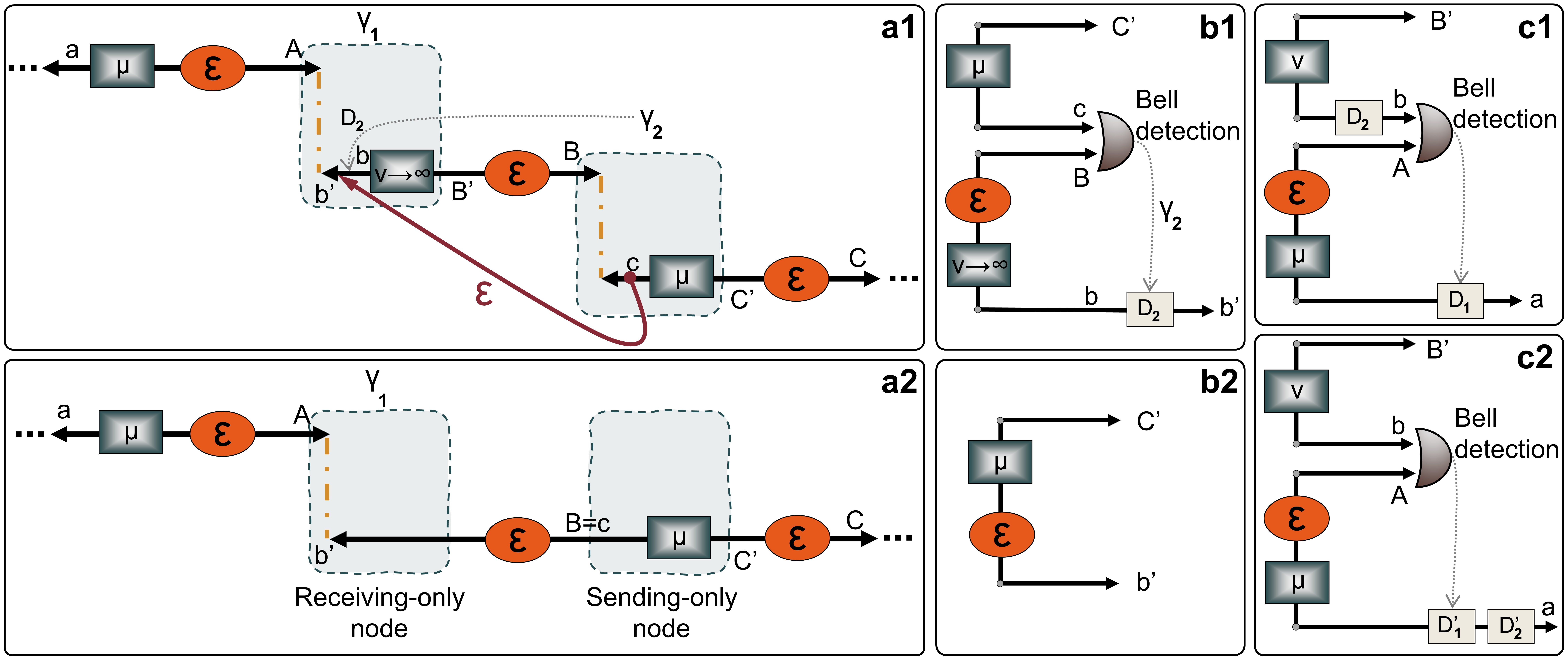}
\caption{{\bf Emulation of specific nodes of a quantum network.} {\bf a1} and {\bf a2,} We emulate receiving-only and sending-only nodes. 
{\bf b1} and {\bf b2,} We sketch the teleportation stretching form of {\bf a1} and {\bf a2}, respectively. 
{\bf c1} and {\bf c2,} We show that all displacement operations can be postponed to one, receiver end (see main text). Here, $\mathcal{E}$ and $\widehat{D}$ represent a thermal-loss channel and single-mode displacement operation, respectively.  
(see main text for explanation). } 
\label{fig:NodeEmulation}
\end{figure*}

\smallskip
\textbf{Emulation of sending- and receiving-only relays}.
It is conceivable that a node in a quantum network is exploited to only send/share or only receive/measure quantum signals. 
In order to keep our study as general as possible, especially when it comes to parameter estimation, we shall simulate such specific relays that include either a relay with some outcome $\gamma$ or a source with some variance $\mu$ to feed its adjacent relays; see Fig.~\ref{fig:NodeEmulation}a1. 
Assume three single links that are connected via two Bell detection modules. The emulation can be performed by 
(i) applying the second relay on modes $B$ and $c$, which produces the outcome $\gamma_2$, 
(ii) applying a correction/displacement, $\widehat{D}_2$, at the first relay on mode $b$, which subsequently teleports mode $c$ to $b'$, and
(iii) taking the limit $\nu \rightarrow \infty$. 
As sketched in Fig.~\ref{fig:NodeEmulation}a2, we show that the above steps would reduce the two `full' relays, which include a Bell measurement as well as a TMSV source, to a receiving-only and a sending-only relays. 

For convenience, let us describe the situation in terms of the teleportation-stretching technique, developed in~\cite{Pirandola:PLOB2017}, as shown in Fig.~\ref{fig:NodeEmulation}b1 and b2. We shall show that in both cases, after taking the limit $\nu \rightarrow \infty$, the resultant CM for modes $b'C'$ is the same. By assuming that $\mathcal{E}$ is a thermal-loss with transmissivity $\eta$ and noise at channel output $\xi$, we see that the scheme in Fig.~\ref{fig:NodeEmulation}b2 gives
\begin{align}
\label{eq:CMEmula}
\mathbf{V}_{b'C'}^{\rm b2}=
\left(\begin{array}{cc}
\mu \mathbf{I} & \sqrt{\eta(\mu^2-1)} \mathbf{Z} \\
\sqrt{\eta(\mu^2-1)} \mathbf{Z} & \eta\mu +1-\eta +\xi \mathbf{I}
\end{array}\right).
\end{align}
With a bit of math one can show that the execution of the relay in Fig.~\ref{fig:NodeEmulation}b1 gives
\begin{align}
\mathbf{V}_{b'C'}^{\rm b1}= &
\left(\begin{array}{cc}
\mu-\frac{\mu^2-1}{\mu+\nu}  \mathbf{I} &  \frac{\sqrt{\eta (\mu^2-1)(\nu^2-1)}}{\mu+\nu} \mathbf{Z} \\
 \frac{\sqrt{\eta (\mu^2-1)(\nu^2-1)}}{\mu+\nu} \mathbf{Z} &  F \mathbf{I}
\end{array}\right), \\
F= & \frac{\mu(\eta\mu +1-\eta +\xi)+(1-\eta+\xi)+\eta}{\mu+\nu}. \notag
\end{align}
One can then verify that $\mathbf{V}_{b'C'}^{\rm b1}$ equals the CM in Eq.~\eqref{eq:CMEmula} in the limit $\nu \rightarrow \infty$.

In Fig.~\ref{fig:NodeEmulation}c1 and c2, we further verify that corrections based on broadcasted $\gamma_i$'s can be postponed to one end (here to the end-mode `$a$'). We do so by checking the equivalence when the displacement operator $\widehat{D}_2$ can be postponed and performed along with $\widehat{D}_1$. The equivalence can be verified through checking both CMs and mean values. From upper and lower panels in Fig.~\ref{fig:NodeEmulation}c, it is clear that the equality holds for CMs since both scenarios start with the same resources and channels, on which only local displacement operations, which do not change the CMs, are applied. 

For mean values, we start by the fact that initial mean value vector for the four involved modes is zero, i.e., $\overline{\mathbf{x}}_{B'bAa}= \mathbf{0}$. 
Let us start from Fig.~\ref{fig:NodeEmulation}c1. 
The displacement $\gamma_2:=(q_{\gamma_2}+ip_{\gamma_2})/\sqrt{2}$ implies that $\overline{\mathbf{x}}_{B'bAa}'= (0~0~q_{\gamma_2}~p_{\gamma_2}~0~0~0~0)^\mathsf{T}$, which after applying the balanced beam splitter of the Bell detection varies to
\begin{align}
    \overline{\mathbf{x}}_{B'bAa}''= \Big(0~0~\frac{q_{\gamma_2}}{\sqrt{2}}~\frac{p_{\gamma_2}}{\sqrt{2}}~ \frac{-q_{\gamma_2}}{\sqrt{2}}~\frac{-p_{\gamma_2}}{\sqrt{2}}~0~0\Big)^\mathsf{T}.
\end{align}
Next, it can be shown that the execution of homodyne detection modules, with the outcomes $q_{\gamma_1}$ and $p_{\gamma_1}$ that forms  $\gamma_1:=q_{\gamma_1}+ip_{\gamma_1}$, gives the mean value vector for the mode $a$

\begin{align}
\overline{\mathbf{x}}_{a}'''= \Gamma 
\left(\begin{array}{c}
q_{\gamma_2}+\sqrt{2}q_{\gamma_1} \\
p_{\gamma_2}-\sqrt{2}p_{\gamma_1}
\end{array}\right),\text{~with~}
\Gamma:=\frac{\sqrt{\eta(\mu^2-1)}}{\nu+\eta(\mu-1)+1+\xi}.
\end{align}

Then the parties apply the following displacement dependent on the outcomes 
\begin{equation}
    D_1(q_{\gamma_1},p_{\gamma_1})=\Gamma\begin{pmatrix} -\sqrt{2}q_{\gamma_1}\\+ \sqrt{2}p_{\gamma_1}
    \end{pmatrix},
\end{equation}
and obtain the mean of mode $b$ (rescaled by the factor $\Gamma$) 
\begin{equation}
\overline{\mathbf{x}}^{\rm up}_a=\Gamma \begin{pmatrix} q_{\gamma_2} \\ p_{\gamma_2}\end{pmatrix}.
\end{equation}
In Fig.~\ref{fig:NodeEmulation}c2, after applying the Bell detection module, with outcomes $q^\prime_{\gamma_1}$ and $p^{\prime}_{\gamma_1}$, we have that
\begin{equation}
    \overline{\mathbf{x}}_a= \sqrt{2} \Gamma\begin{pmatrix}
    q^\prime_{\gamma_1} \\ p^{\prime}_{\gamma_1}
    \end{pmatrix};
\end{equation}
hence, we apply the displacement $D'_1(q^\prime_{\gamma_1},p^\prime_{\gamma_1})$ to obtain
\begin{equation}
 \overline{\mathbf{x}}_a'=\begin{pmatrix}
 0\\0
 \end{pmatrix}.
\end{equation}
One can show that the mean value vector for the mode $a$
after the displacement 
$D^\prime_2(q^\prime_{\gamma_2},p^\prime_{\gamma_2})$ is given by
\begin{equation}
   \overline{\mathbf{x}}_{a}^{\rm down}=  \begin{pmatrix}
    q^\prime_{\gamma_2} \\ p^\prime_{\gamma_2}
    \end{pmatrix},%=\Gamma \begin{pmatrix}
    % q_{\gamma_2} \\ p_{\gamma_2}
   % \end{pmatrix}
\end{equation}
whose entries can be tuned so that $\overline{\mathbf{x}}_{a}^{\rm down}=\overline{\mathbf{x}}_{a}^{\rm up}$. 
% is given by
% \begin{align}
% \overline{\mathbf{x}}_{a}^{\rm down}= \Gamma
% \left(\begin{array}{c}
% q^\prime_{\gamma_1} \\
% p^\prime_{\gamma_1}
% \end{array}\right) +
% \left(\begin{array}{c}
% q_{\gamma_2}' \\
% p_{\gamma_2}'
% \end{array}\right). 
% \end{align}
% in which one can tune 
%\begin{equation}
%    \begin{pmatrix}
%    q^\prime_{\gamma_2} \\ p^\prime_{\gamma_2}\end{pmatrix}=\Gamma \begin{pmatrix}
%     q_{\gamma_2} \\ p_{\gamma_2}
%    \end{pmatrix}.
%\end{equation}
%=\Gamma \begin{pmatrix}
    % q_{\gamma_2} \\ p_{\gamma_2}
   % \end{pmatrix}
%$\gamma_2'$ such that $\overline{\mathbf{x}}_{a}^{\rm down}=\overline{\mathbf{x}}_{a}^{\rm up}$. 

Note that we can define a ``network number,'' $S_{\rm net}$, which tells us how a generic network is different from a fully designed network whose nodes are all sending-receiving. Precisely, the number $S_{\rm net}$ accounts for the number of only-receiving plus only-sending nodes. A fully designed network then has $S_{\rm net}=0$. Note also that, as described in Fig.~\ref{fig:NodeEmulation}, such nodes always appear in pairs, such that $S_{\rm net}$ is an even number, the reason being directionality of the generated signals as well as network’s symmetry. In fact, like an entanglement source that feeds the two ends of a single link, the function of the network is to distribute entanglement {\em towards} both far-ends. Installing nodes that result in $S_{\rm net}$ being an odd number, would break the directionality, and the symmetry, which therefore breaks off the links of the network.

\smallskip
{\bf Parameter estimation.}   
The outcomes of the relays are bi-dimensional Gaussian variables $\gamma_i=(q_{\gamma_i},p_{\gamma_i})^T$, which are taken into account by Alice and Bob to post-process their local variables. Let us focus on the $q$-quadrature for the next derivations since equivalent steps hold for the $p$-variable. To simplify the derivations, we in fact assume that $q$ and $p$ quadratures are not mixed by the eavesdropper so that they can be treated as independent variables. (This is a reasonable protocol assumption; extension is just a matter of technicalities). 
 
In this work we assume that both Alice and Bob apply heterodyne measurements on the end-to-end modes $A'$ and $B'$ with outcomes $z_A:=(q_A,p_A)$ and $z_B:=(q_B,p_B)$, respectively, to establish a secure key. 
% Equation~\eqref{cond_measCM} would change if, e.g., Bob performs a homodyne detection~\cite{Weedbrook:GaussQI2012}.
By assuming that the relays work properly and that the quadratures  follow a normal distribution, we can write the variables that build the raw key for Alice and Bob, respectively, as 
\begin{align}
q_x=& q_A-\sum^{M-1}_{i=1}u_iq_{\gamma_i}, \label{eq:correction vars1} \\
q_y=& q_B- \sum^{M-1}_{i=1}v_iq_{\gamma_i},\label{eq:correction vars2}
\end{align}
where $u_i$'s and $v_i$'s are real numbers.\footnote{In a prepare and measure variant, where Alice prepares coherent states, she generates variable $\bar{z}_A=\left(\sqrt{\frac{\mu-1}{\mu+1}}q_A,\sqrt{\frac{\mu-1}{\mu+1}}p_A\right)$, with $\mu=\sigma^2_A-1$, where $\sigma^2_A$ is the variance of the Gaussian modulation of $\bar{z}_A$. Hence, before applying Eqs.~\eqref{eq:correction vars1} and \eqref{eq:correction vars2}, one needs to apply a transformation, $\mathbf{L}=\sqrt{\frac{\mu-1}{\mu+1}}\mathbf{I}$, on $\bar{z}_A$ in order to obtain $z_A$.}
%Here the variables $q_A$ and $q_B$ are Alice's 
%preparation variable 
%and Bob's measurement outcome respectively, while $q_x$ and $q_y$ are the variables that build the raw key.
For security reasons, we require $q_x$ and $q_y$ to be uncorrelated with the public variables $q_{\gamma_i}$'s that are known to Eve, i.e., $\langle q_x q_{\gamma_i} \rangle = 0$, which imposes the following constraints
\begin{align}
 \langle q_A q_{\gamma_i} \rangle=\sum^{M-1}_{k=1} u_k \langle q_{\gamma_i}q_{\gamma_k} \rangle,
\label{eq:y_gamma}
\end{align}
from which one can calculate the weights $u_i$'s (similar relations hold for $q_y$,  $q_B$ and $v_i$'s).

Now, let us consider and study the sampled data $[q_A]_j$ and $[q_{\gamma_i}]_j$, for $j=1, \dots, N$, associated with variables $q_A$ and $q_{\gamma_i}$, respectively.
From these, Alice can calculate the corresponding maximum likelihood estimators % \SP{what is $p_s$ below?}
\begin{align}
\widehat{\langle q_A q_{\gamma_i} \rangle} &= N^{-1} \sum_{j=1}^N [q_A]_j[q_{\gamma_i}]_j, \\
\widehat{\langle q_{\gamma_i} q_{\gamma_k} \rangle} &= N^{-1} \sum_{j=1}^N [q_{\gamma_i}]_j[q_{\gamma_k}]_j.
\end{align}
%directly by manipulating $p_{\rm s}N$ samples, $[]_j$'s, i.e., all the data, which renders the uncertainty of these values negligible.
Next, to obtain values of the weights $u_i$'s, she replaces these values in the set of $M$ equalities in Eq.~(\ref{eq:y_gamma}). She then continues with calculating $[q_x]_j$ by replacing the $u_i$'s, and the data points $[q_A]_j$ and $[q_{\gamma_i}]_j$, in Eq.~\eqref{eq:correction vars1}. Indeed, Bob obtains similar relations for $q_y$ and $v_i$'s.

At this stage the parties are in a position to compute the {classical} CM associated to their post-processed data
\begin{align}
\widehat{\boldsymbol{\Sigma}}=\begin{pmatrix}
\widehat{\mathbf{V}}_x&\widehat{\mathbf{C}}_{xy}\\
\widehat{\mathbf{C}}_{xy}&\widehat{\mathbf{V}}_y
\end{pmatrix},  \label{SP3}
\end{align}
where $\widehat{\mathbf{V}}_x=\text{diag}\big(\widehat{\langle q_x^2\rangle},\widehat{\langle p_x^2\rangle}\big)$,  
$\widehat{\mathbf{V}}_y=\text{diag}\big(\widehat{\langle q_y^2\rangle},\widehat{\langle p_y^2\rangle}\big)$, and 
$\widehat{\mathbf{C}}_{xy}=\text{diag}\big(\widehat{\langle q_xq_y\rangle},\widehat{\langle p_xp_y\rangle}\big)
$.

%\SP{****3. Do not understand this notation. Perhaps you wanted to write $\widehat{\mathbf{V}}_x= \mathrm{diag} \big(\widehat{\langle q_x^2\rangle},\widehat{\langle p_x^2\rangle}\big)$ and so on?****}

For the $q$-quadrature we have that (not to mention that the parties repeat the same process for the $p$-quadrature)
\begin{align}
\widehat{\langle q_x^2 \rangle}&=m_\text{pe}^{-1}\sum_{j=1}^{m_\text{pe}}[q_x]_j^2, \\
\widehat{\langle q_y^2 \rangle}&=m_\text{pe}^{-1}\sum_{j=1}^{m_\text{pe}}[q_y]_j^2, \\
\widehat{\langle q_xq_y \rangle}&=m_\text{pe}^{-1} \sum_{j=1}^{m_\text{pe}}[q_x]_j[q_y]_j,
\end{align}
when $m_{\rm pe}$ is the number of signals sacrificed for PE. 

Note that, in principle, the parties can locally calculate the values from the estimators $\widehat{\mathbf{V}}_x$ and  
$\widehat{\mathbf{V}}_y$ using $N$ data points while $\widehat{\langle q_xq_y \rangle}$ demands sharing $m_{\rm pe}$ data points through the public classical channel. These data can be easily acquired by Eve and thus must not contribute to the key generation. In general, the parties optimize the amount of shared data, $m_{\rm pe}$, so as to limit the uncertainty of terms such as $\widehat{\langle q_x q_y \rangle}$ while still keeping as many samples as possible for the secret key. 

The parties can compute an interval, with confidence $1-\varepsilon_{\rm pe}$, for the estimated CM from which they derive the worst-case scenario CM, i.e., the CM that minimizes the key rate according to the sampled data with a probability larger than $1-\varepsilon_{\rm pe}$. 
This CM is given by
\begin{align}\label{eq:CMWorstCase}
\boldsymbol{\Sigma}_{\rm wc}=\widehat{\boldsymbol{\Sigma}}+\sqrt{\frac{4\kappa}{ m_{\rm pe}}}
\begin{pmatrix}
\widehat{\mathbf{V}}_x & -\frac{\widehat{\mathbf{V}}_x+\widehat{\mathbf{V}}_y}{2}\mathbf{Z}\\
-\frac{\widehat{\mathbf{V}}_x+\widehat{\mathbf{V}}_y}{2}\mathbf{Z}&\widehat{\mathbf{V}}_y
\end{pmatrix},
\end{align}
with $\kappa = \text{ln}(8\varepsilon_{\rm pe}^{-1})$. This is calculated by using suitable tail bounds for the chi-squared distribution (see Methods). It is valid for any CM of two correlated systems even if the entries are given theoretically via a model, e.g., $y=\sqrt{\tau}x+\epsilon$, with scale factor $\sqrt{\tau}$ and variance $\sigma^2_\epsilon$ for the normal variable $\epsilon$.
% Then the parties calculate  worst-case scenario {quantum} CM through the transformation 
% \begin{align}
% % eq:transformation_three_links CCM->QCM
% \label{CM:data}
%     \mathbf{V}_{\rm wc}=\mathbf{U}\boldsymbol{\Sigma}_{\rm wc}\mathbf{U}-\mathbf{I},
% \end{align}
% where $\mathbf{U}=\sqrt{\frac{\mu+1}{\mu-1}}\mathbf{Z}$, with $\mu=\sigma_A^2+1$. 

\smallskip
{\bf Asymptotic key rate.}
We define the asymptotic secret key rate of sharing a key between two arbitrary users of a quantum network based on the Devetak-Winter rate~\cite{Devetak:RateDef2005}
\begin{align}
\label{eq:rate}
K=\beta I(z_A:z_B|\{\gamma_i\})-\chi(E:z_r|\{\gamma_i\}), 
\end{align}
where $I(z_A:z_B|\{\gamma_i\})$ is the mutual information between $z_A$ and $z_B$ %(the variables of the parties who reconcile to build the secret key)
and $S(E:z_r)$ is the Holevo information between Eve's system and the reconciliation variable $z_r$, with $r=A(B)$ indicating direct (reverse) reconciliation. In this work, we focus on the reverse reconciliation $r=B$. 
By definition, we have 
\begin{align}
\chi(E:z_B|\{\gamma_i\})=S(E|\{\gamma_i\})-S(E|z_B\{\gamma_i\}),
\end{align}
where 
\begin{align}
    S(E|\{\gamma_i\})=-\text{tr}\big(\rho_{E|\{\gamma_i\}}\log_2\rho_{E|\{\gamma_i\}}\big)
\end{align}
is the von Neumann entropy of Eve's state, $\rho_E$ (conditioned on the knowledge of the $\gamma$'s), and 
\begin{align}
S(E|z_B\{\gamma_i\})=& \int \mathrm{d}z_B p(z_B|\{\gamma_i\})\notag\\ &\times\big[-\text{tr}\big(\rho_{E|z_B\{\gamma_i\}}\log_2\rho_{E|z_B\{\gamma_i\}}\big)\big], 
\end{align}
where $\rho_{E|z_B\gamma_i}$ is the state conditioned on Bob's variable $z_B$ (after the $\gamma$'s).

% In our study we consider the entanglement-based representation for Alice, meaning that she has a TMSV state and she heterodynes one of the modes with outcome 
% \begin{equation}\label{eq:rescaling}
% z_{A}=(\sqrt{{\mu+1}{\mu-1}}q_A,\sqrt{{\mu+1}{\mu-1}}p_A)^T.
% \end{equation}
% This outcome will be distributed according to a bi-variate Gaussian with variance $\mu+1$. This is equivalent to creating an ensemble of Gaussian-modulated coherent states on the other, un-measured mode sent through the quantum network. The created coherent-state ensemble has a mean value $\bar{z}_{A}=(q_{A},p_A)$  which is Gaussian distributed with variance $\sigma_{A}^{2}=\mu -1$. At the output of the network, Bob heterodynes the incoming mode with outcome $z_{B}=(q_B,p_B)^T$. However, some protocols demand entanglement for all their steps (see~\cite{Ghalaii:QRs}), i.e., the preparation stage is based on a TMSV state, where one of the modes is measured so that the other (traveling one) is projected to a coherent state. Then one may write directly
% \begin{equation}\label{eq:direct}
% z_{A}=(q_A,p_A)^T,
% \end{equation}
% where $q_A$ and $p_A$ are the outcomes of a measurement.

With this in mind, the parties neither know the explicit description of Eve's system nor how she interacts with the links. However, by assuming that Eve purifies the system between Alice and Bob, such that $\rho_{ABE|\{\gamma_i\}}$ is a pure state, it holds that $S(E|\{\gamma_i\})=S(AB|\{\gamma_i\})$ and $S(E|z_B\{\gamma_i\})= S(A|z_B\{\gamma_i\})$~\cite{ArakiLieb:1970}, where the later equality also exploits the fact that Bob performs a rank-1 measurement (like heterodyne detection) therefore projecting the global pure state $\rho_{ABE|\{\gamma_i\}}$ into a reduced pure state $\rho_{AB|z_B\{\gamma_i\}}$.
Since the state $\rho_{AB|\{\gamma_i\}}$ is Gaussian, it is characterized by its CM, $\mathbf{V}_{AB|\{\gamma_i\}}$. 
In practice, this can be estimated by the worst-case quantum CM, $\mathbf{V}_{\rm wc}$ (compatible with the classical data shared by the parties)
% \begin{align}
% % eq:transformation_three_links CCM->QCM
% \label{CM:data}
% \mathbf{V}_{AB|\{\gamma_i\}}:= \mathbf{V}_{\rm wc} = \mathbf{L}^{-1}\boldsymbol{\Sigma}_{\rm wc}(\mathbf{L}^{-1})^\mathsf{T}-\mathbf{I}\oplus\mathbf{I},
% \end{align}
\begin{align}
% eq:transformation_three_links CCM->QCM
\label{CM:data}
\mathbf{V}_{AB|\{\gamma_i\}}\simeq \mathbf{V}_{\rm wc} = \boldsymbol{\Sigma}_{\rm wc}-\mathbf{I}\oplus\mathbf{I}.
\end{align}
% with $\mathbf{L}=\left(\sqrt{(\mu-1)/(\mu+1)}\mathbf{I}\right)\oplus\mathbf{I}$ expressing the rescaling factors from the P\&M to EB representation (see Eq.~\eqref{eq:rescaling}) and $\mu=\sigma_A^2+1$. The quantity $\sigma_A^2$ can be estimated   with negligible variance since all the data $[q_A]_i$ (only from Alice's preparation) can be used. Otherwise, when the preparation step is taking place through a measurement as in Eq.~\eqref{eq:direct}, the parties
% obtain the CM in Eq.~\eqref{CM:data} but with $\mathbf{L}$ being the identity matrix.% , and $\mathbf{I}\otimes\mathbf{I}$ is the four dimensional identity matrix. 
%A similar expression holds for the state $\rho_{A|z_B\{\gamma_i\}}$; hence, one can find
By setting 
\begin{align}\label{cond_cm}
\mathbf{V}_{AB|\{\gamma_i\}}:=\begin{pmatrix}
\mathbf{V}_{A|\{\gamma_i\}}&\mathbf{C}_{AB|\{\gamma_i\}}\\
\mathbf{C}_{AB|\{\gamma_i\}}&\mathbf{V}_{B|\{\gamma_i\}}
\end{pmatrix}, 
\end{align}
we have that the conditional CM after Bob's heterodyne is given by 
\begin{align}\label{cond_measCM}
\mathbf{V}_{A|z_B\{\gamma_i\}}=\mathbf{V}_{A|\{\gamma_i\}}-\mathbf{C}_{AB|\{\gamma_i\}}^\top\left[\mathbf{V}_{B|\{\gamma_i\}}+\mathbf{I}\right]^{-1}\mathbf{C}_{AB|\{\gamma_i\}}. \end{align}

Next, from the symplectic spectra $\mathbf{\nu}_{AB|\{\gamma_i\}}$ and  $\mathbf{\nu}_{A|z_B\{\gamma_i\}}$, of $\mathbf{V}_{AB|\{\gamma_i\}}$ and  $\mathbf{V}_{A|z_B\{\gamma_i\}}$, we compute the Holevo information
\begin{align}
\chi(E:z_B|\gamma_i)=\sum_l h([\mathbf{\nu}_{AB|\{\gamma_i\}}]_l)-\sum_k h([\mathbf{\nu}_{A|z_B\{\gamma_i\}}]_k), 
\end{align}
where $h(x)=\frac{x+1}{2}\log_2\frac{x+1}{2}- \frac{x-1}{2}\log_2\frac{x-1}{2}$.
In addition, the mutual information is given by~\cite{Pirandola:MDI15} \begin{align}\label{mutual_info}
     I(z_A:z_B|\{\gamma_i\})=\frac{1}{2}\log_2\frac{1+\det\mathbf{V}_{A|\{\gamma_i\}}+\text{tr} \mathbf{V}_{A|\{\gamma_i\}}}{1+\det \mathbf{V}_{A|z_B\{\gamma_i\}}+\text{tr} \mathbf{V}_{A|z_B\{\gamma_i\}}}.
\end{align}

Therefore, the parties have calculated a modified asymptotic key rate that encompasses the worst-case scenario given the sampled data. This is correct up to an error $\varepsilon_{\rm pe}$ and is calculated from Alice's and Bob's remote, shared, data that account for the relay outputs $\gamma_i$'s without any assumption on the structure of the intermediate channels. To put it more precisely, the rate should be re-scaled in a way to account for the number of uses sacrificed for parameter estimation. We discuss this and other aspects in detail shortly. 

Let us also remark that in all the derivation above, we assume that the conditioning associated with the $\gamma_i$'s create the same conditional CM for the shared state regardless of the actual values of $\gamma_i$'s.  
This makes sense only under the Gaussian assumptions for the network, but this is still true even in the presence of NLAs, where the network is conditionally Gaussian.

\smallskip
{\bf Composable finite-size key rate.}
The security of Gaussian quantum networks can be further extended by considering finite-size correction terms dependent on small failure probabilities of different processes of the protocol. 
%, such as the error correction (EC) and the privacy amplification (PA), which describes a practical scenario away from the ideal case where all the processes are considered successful. 
%In this case the composable secret key rate is given by~\cite{Pirandola:FS2021}
%\begin{align} \label{comprate}
%K \geq & \frac{np_{\rm ec}}{N} \Big( R_{\rm pe}- \frac{\Delta_{\rm aep}}{\sqrt{n}} + \frac{\Theta}{n} \Big),
%\end{align} where 
%\begin{align}
 %   \Delta_{\rm aep}:= & 4\log_2(\sqrt{d}+2) \sqrt{\log_2(18 p_{\rm ec}^{-2} \varepsilon_{\rm s}^{-4} )} , \\
 %   \Theta:= & \log_2\big[p_{\rm ec}(1-\varepsilon_{\rm s}^2/3)\big] + 2\log_2(\sqrt{2}\varepsilon_{\rm h}), 
%\end{align}
%and the small failure probabilities, $\varepsilon_\text{cor}$ for EC, $\varepsilon_{\rm pe}$ for PE, and $\varepsilon_\text{s}$, $\varepsilon_\text{h}$ for PA, are composed ito one security parameter $\varepsilon=\varepsilon_{\rm cor} +  \varepsilon_{\rm s} +  \varepsilon_{\rm h} + 2 p_{\rm ec}\varepsilon_{\rm pe}$. Note that $p_{\rm ec}$ is the probability of a block of channel uses to pass the verification step after EC while $d$ is the discretization parameter.
Over a chosen route of the network, Alice and Bob would share the following classical-quantum state between themselves and Eve, who is assumed to perform a collective Gaussian attack,
\begin{align}
\rho_{ABE}= \sum_{k,l} p(k,l) |k\rangle_A \langle k| \otimes  |l\rangle_B \langle l| \otimes \rho_E(k,l),
\end{align}
where $E\equiv E_1E_2\dots E_M$ are Eve's systems; see Fig.~\ref{fig:qchain}a. Thus, at the end of the error correction, Alice and Bob possess correlated discretized variables $k^n$ and $l^n$ respectively associated with $\rho_{ABE}^{{\otimes n}}$. 

As we discussed, the end-to-end CM, either built from sampled data or given by means of a proper model, would suffice to derive the secret key rate or suitable bounds by using the notions of coherent information and reverse coherent information of bosonic channels~\cite{GarciaPatron:RCI2009,Pirandola:RCI2009}, as well as the relative entropy of entanglement~\cite{Vedral:REE2002}. 
Since in a real-world scenario the parties exchange only a finite number of signal states, here the focus is put on composable finite-size analysis, which has become the touchstone for QKD security, rather than the ultimate bounds. 
The security of a QKD protocol is desired to be composable, i.e., the protocol must not be distinguished from an ideal protocol which is secure by construction~\cite{Pirandola:AQCrypt}. Mathematically, a composable security proof can be provided by incorporating proper error parameters, $\varepsilon$'s, for each segment of the protocol, namely, error correction (EC), privacy amplification (PA), smoothing, and hashing~\cite{Tomamichel:2012,Furrer:FSize2012}.

We assume that a total number of $N$ Gaussian signals are measured by Alice and Bob. An amount $n$ of these would be used for key extraction, while the rest $m_{\rm pe}=N-n$ are left for PE, i.e., the evaluation of the CM. Upon successful completion of the EC procedure, with probability $p_{\rm ec}$, the composable finite-size secret key rate is given by~\cite{Pirandola:FS2021}
\begin{align}
\label{comprate}
K \geq & \frac{np_{\rm ec}}{N} \Big( K_{\rm pe}- \frac{\Delta_{\rm aep}}{\sqrt{n}} + \frac{\Theta}{n} \Big),
\end{align}
where the higher-order terms read 
\begin{align}
    \Delta_{\rm aep}:= & 4\log_2(\sqrt{d}+2) \sqrt{\log_2(18 p_{\rm ec}^{-2} \varepsilon_{\rm s}^{-4} )} , \\
    \Theta:= & \log_2\Big[p_{\rm ec}(1-\frac{\varepsilon_{\rm s}^2}{3})\Big] + 2\log_2(\sqrt{2}\varepsilon_{\rm h}),
\end{align}
The above equation is valid for a protocol with overall security $\varepsilon= \varepsilon_{\rm cor} + \varepsilon_{\rm s} +  \varepsilon_{\rm h} + p_{\rm ec}\varepsilon_{\rm pe}$, where $\varepsilon_{\rm pe}$
is the total error probability associated with PE. Assuming reverse reconciliation, the hash comparison stage of the finite-key analysis requires Bob sending $\ceil[\big]{\log_2(1-\varepsilon_{\rm cor})}$ bits to Alice for some proper $\varepsilon_{\rm cor}$ (called $\varepsilon_{\rm cor}$-correctness) and bounds the probability that Alice's and Bob's sequences are different even if their hashes coincide. $\varepsilon_{\rm h(s)}$ is the hashing (smoothing) parameter. Conveniently one can also define the frame error rate $\textsc{FER}=1-p_{\rm ec}$. 
It is also assumed that by using an analog-to-digital conversion, each continuous-variable symbol is encoded with $d$ bits of precision.

The value of $K_{\rm pe}$ in Eq.~\eqref{comprate} can be computed in different ways depending on the level of reliability. 
In practice, one would use the sampled data to compute $K_{\rm pe}$ using Eq.~\eqref{eq:rate} and the worst-case CM shared by the end-users. Remarkably this is practically the most appropriate choice in the case of multi-hop quantum networks with untrusted relays. 
In the presence of a conditionally Gaussian network, the rate in Eq.~\eqref{comprate} modifies by setting $m\rightarrow mp_{\rm s}$ where $p_{\rm s}$ is the probability of successful post-selection. 
As an example, in the following, we study a quantum repeater chain and compute the composable finite-size key rate considering the worst-case parameters for the end-to-end shared CM. 

%Back to the estimation of the worst-case parameters, by revealing $m$ pairs of corresponding data, i.e., $[x]_i$ and $[y]_i$, Alice and Bob can build an estimator $\widehat T$ of the square root of transmissivity $T=\sqrt{\eta}$, that is $\widehat T:=m^{-1}\sigma_x^{-2} \sum_{i=1}^{m} x_iy_i$, with variance $\text{Var}(\widehat T)=m^{-1}(2\eta + \sigma_x^{-2}\sigma_z^2)$, where $\sigma_x^2=\sum_{i=1}^{m} x_i^2\simeq \mu-1$. Then, the estimator for transmissivity is $\widehat \eta=(\widehat T)^2$, with variance $\text{Var}(\widehat \eta)=4m^{-1}\eta^2 \big(2+ \eta^{-1} \sigma_x^{-2} \sigma_z^2 \big)+\mathcal{O}(m^{-2})$. Similarly, Alice and Bob can construct the estimator for  $\bar n$, that is, $\widehat{\bar n}:=(\widehat{\sigma_z^2}-1)/2$, with variance $\text{Var}(\widehat{\bar n})=\sigma_z^4/(2m)$. Here, $\widehat{\sigma_z^2}=m^{-1}\sum_{i=1}^{m} z_i^2$ is the the estimator for the variance of the thermal noise $\sigma_z^2$. 

%Next, by assuming a certain number $w$ of confidence of intervals, Alice and Bob compute the worst-case estimators up to some probability of error $\varepsilon_{\rm pe}=\big[1-\text{erf}(w/\sqrt{2}) \big]/2$, i.e., \begin{align}
%    \eta_{\rm wc}= \eta - 2w\sqrt{\frac{2\eta^2 +\eta \sigma_x^{-2} \sigma_z^2 }{m}}, ~
%    \bar n_{\rm wc}=  \bar n + w \frac{\sigma_z^2}{\sqrt{2m}}. 
%\end{align}

\smallskip 
{\bf Numerical results.}
As we mentioned earlier, a route between two nodes in a quantum network can be seen as a chain of quantum links. We here apply our general techniques for quantum networks to study a quantum chain of {\em identical} links and generally-untrusted stations. We note that this is a mere example and that our methodology is generic that can be applied to any chain. Subsequently, by assuming the illustration in Fig.~\ref{fig:qchain}, we find the end-to-end CM and compute the composable finite-size key rate. 

Let us assume the chain is made of $M=2^m$ identical links (we call $m$ the repeater depth), each described by a standard CM
\begin{align}
\mathbf{V}_{0} =
\left(\begin{array}{cc}
\mathsf{a}_0 \mathbf{I} & \mathsf{c}_0 \mathbf{Z} \\
\mathsf{c}_0 \mathbf{Z} &  \mathsf{b}_0 \mathbf{I}
\end{array}\right).
\end{align}
For a typical link (without an NLA), with a TMSV source $\mu$, channel loss $\eta$ and excess noise $\epsilon$ (referred to the channel's input), we have that 
$\mathsf{a}_0=\mu$, $\mathsf{b}_0=\eta\mu+1-\eta+\eta\epsilon$, and $\mathsf{c}_0=\sqrt{\eta(\mu^2-1)}$.
By using similar techniques introduced in~\cite{Ghalaii:QRs}, the end-to-end CM between Alice and Bob, in the case of non-ideal Bell measurements, is found to have the standard form 
\begin{align}
\label{CM:recursive}
\mathbf{V}_{AB} =
\left(\begin{array}{cc}
\mathsf{a}_m \mathbf{I} & \mathsf{c}_m \mathbf{Z} \\
\mathsf{c}_m \mathbf{Z} &  \mathsf{b}_m \mathbf{I}
\end{array}\right),
\end{align}
with the following parameters 
\begin{align}
\begin{cases}
\mathsf{a}_m=\mathsf{a}_{m-1} - \frac{\eta_{\rm B}\mathsf{c}_{m-1}^2 }{\eta_{\rm B}(\mathsf{a}_{m-1} + \mathsf{b}_{m-1})+1-\eta_{\rm B} }, \\
\mathsf{b}_m= \mathsf{b}_{m-1} - \frac{\eta_{\rm B}\mathsf{c}_{m-1}^2 }{\eta_{\rm B}(\mathsf{a}_{m-1} + \mathsf{b}_{m-1})+1-\eta_{\rm B} },  \\
\mathsf{c}_m= \frac{\eta_{\rm B}\mathsf{c}_{m-1}^2 }{\eta_{\rm B}(\mathsf{a}_{m-1} + \mathsf{b}_{m-1})+1-\eta_{\rm B} }. 
\end{cases}
\end{align}
As expected, for $\eta_{\rm B}=1$ the above equations reduce to the previous results in~\cite{Ghalaii:QRs}.  
Next, we can apply the formula for finite-size key rate, given in Eq.~\eqref{comprate}.  

\begin{figure}[t]
\vspace{+.1cm}
\includegraphics[scale=0.55]{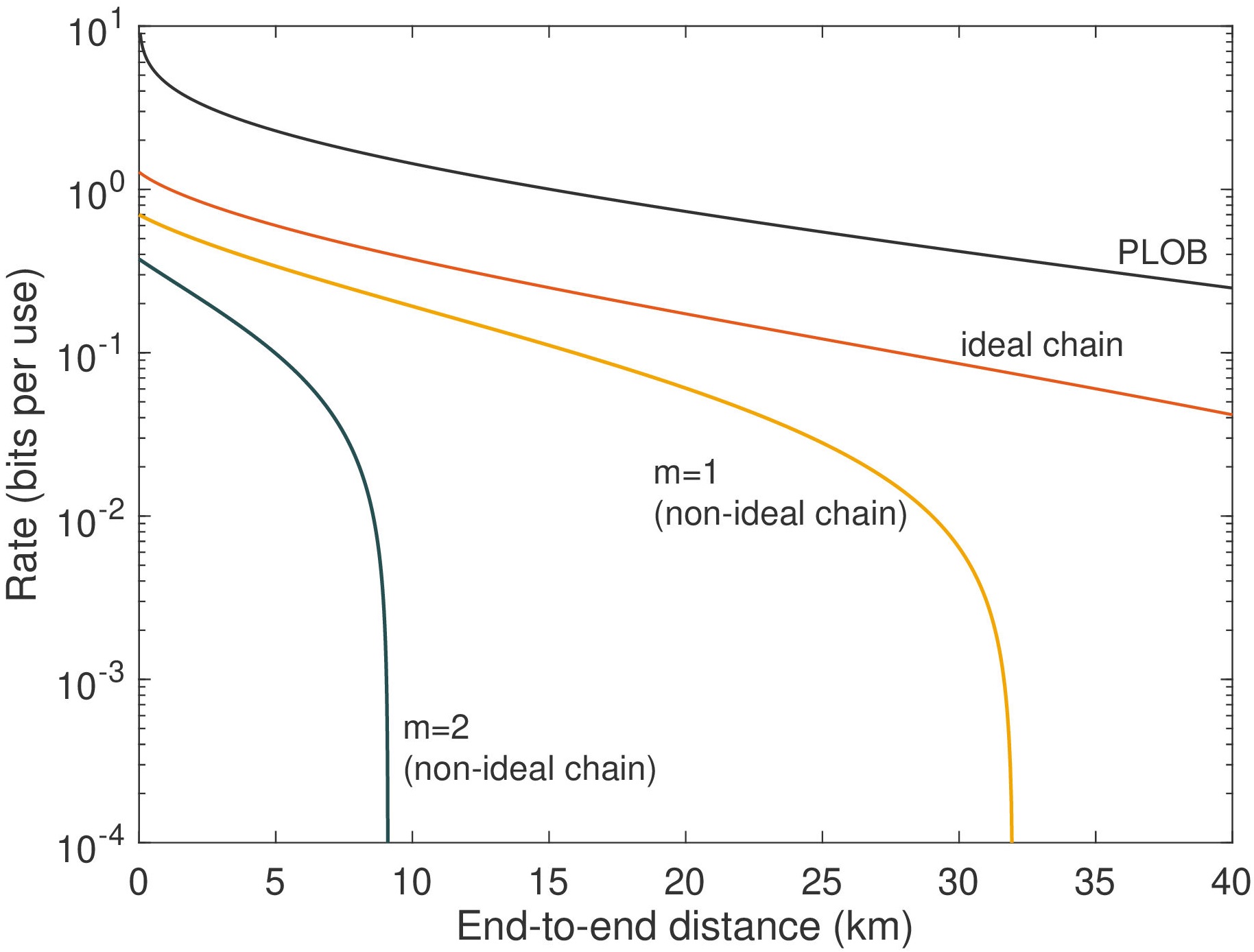}
\caption{{\bf Composable key rate per chain use.} We consider a heterodyne-based CV-QKD protocol implemented over a quantum chain with depths $m=1$ and $m=2$. Here, we assume non-ideal Bell detection modules with $\eta_{\rm B}=0.95$ and excess noise $\epsilon=0.005$~SNU for each single link. 
Other parameters are $\beta=0.98$, $N=10^{10}$, $m_{\rm pe}=0.1N$, $d=2^5$, $\textsc{FER}=0.1$, $\varepsilon_{\rm s}=\varepsilon_{\rm h}=\varepsilon_{\rm pe}=10^{-10}$, $w=6.34$ and $\varepsilon=4.5\times 10^{-10}$. Rates are compared with an ideal chain (with $\eta_{\rm B}=1$, $\epsilon=0$ and $\beta=1$) and the repeaterless capacity, i.e., PLOB bound~\cite{Pirandola:PLOB2017}.}
\label{fig:noNLAchain}
\end{figure}

In Fig.~\ref{fig:noNLAchain}, we plot the secret key rate versus the overall distance between Alice and Bob. Assuming the CV\:QKD protocol with heterodyne detection, we compute $K_{\rm pe}$ for the worst-case scenario CM. The links are thermal-loss channels, which we simulate by considering optical fibres with the loss factor of 0.2~dB/km and noise parameter $\epsilon$. 
Here, we choose an initial modulation at the input of each link, $\mu$, such that the maximum distance is achieved. 
It was observed that the composable rate is highly sensitive to the relay loss, $\eta_{\rm B}$, as well as channel excess noise, $\epsilon$. This can be seen in Fig.~\ref{fig:noNLAchain} where we compare the rates for $m=1$ and $m=2$ with that of ideal chain, with $\eta_{\rm B}=1$ and $\epsilon=0$. 

It is known that Gaussian-only nodes cannot act as quantum repeaters~\cite{Niset:2009,Namiki:2014}. Expectedly, Fig.~\ref{fig:noNLAchain} verifies that the end-to-end rate cannot reach/beat the repeaterless PLOB limit. This is because, in our example, the relays are Gaussian operation and as such they cannot do so. That being said, we emphasise that references~\cite{Niset:2009,Namiki:2014} are more about entanglement distribution than QKD. The quantum repeater chain in our paper has an element of non-Gaussianity in the concept of being post-selectively Gaussian, e.g., via NLAs. 

One can also compare a part of our results to the well-studied measurement-device-independent (MDI) QKD protocols~\cite{Pirandola:CVMDI2015}. In the case where $m=1$ our chain includes two links and one intermediate node, which very much resembles a MDI setup. It is known that the so-called symmetric MDI, wherein the links are identical and the node sits exactly at the middle, is poor in delivering a secret key at long distances, especially for non-zero excess noise and non-ideal relay~\cite{Papanastasiou:FS-MDI2017}. Whereas an asymmetric MDI, wherein the node is closer to one end, can reach longer distances. In our example, we assumed identical links and as such, comparing with symmetric MDI, we do not expect to reach longer distances.

\smallskip 
Now let us revamp the quantum chain to design a quantum repeater. 
Considering the class of continuous-variable quantum repeaters~\cite{Dias:CVQR2017,Dias:CVQR2020,Furrer:CVQR2018,Seshadreesan:QSCVQR2020,Dias:ComparingQRs2019,Ghalaii:QRs}, several proposals have been suggested to increase the reach of single-link CV\:QKD protocols, e.g., by utilizing NLAs~\cite{Blandino:idealNLA2012}, which nevertheless can improve the secure distance for only a few tens of kilometres~\cite{Blandino:idealNLA2012,Ghalaii:JSTQE2020,Ghalaii:JSAC2020}. One idea is that a quantum repeater can essentially be built by a concatenation of such NLA-improved links. Key elements of any repeater chain include entanglement distribution, entanglement distillation or purification, and entanglement swapping. An NLA-based quantum repeater uses TMSV sources as an entanglement distribution source and CV Bell measurements as entanglement swapper device.
Other components such as quantum memories ~\cite{Lvovsky:OptQMs2009,Simon:QMReview2010} can help to improve the performance of quantum repeaters, though they are not essential~\cite{Munro:QCwithoutQR2012,Azuma:All-photonicQR2015,Winnel:2021}. But due to the non-deterministic nature of NLAs, using QMs in the structure of amplifier-based repeaters seems indispensable.

%As we mentioned earlier, a route between two nodes in a quantum network can be seen as a chain of quantum links. By adding auxiliary components such as NLAs to such links, we can turn the route into an effective repeater chain beating the PLOB bound. We here apply our general techniques for quantum networks to study a quantum repeater chain of identical links and generally-untrusted stations, subsequently find the end-to-end CM, and finally compute the composable finite-size key rate. We assume that the chain is arranged as illustrated in Fig.~\ref{fig:qchain}, especially that ideal NLAs, with gain $g$, and TMSV sources, with variance $\mu$, are in working order. 

% By using similar tools, we recently reported that an NLA-based quantum repeater can outdistance the PLOB bound and, furthermore, approach the ultimate end-to-end quantum repeater capacities~\cite{Ghalaii:QRs}. There we allowed for an infinite number of signal exchange and, hence, found the so-called asymptotic secret key rate. Within our illustration, we here present composable security analysis of quantum repeaters. Further, we also relax the assumption of an ideal Bell measurement, i.e., we model non-ideal Bell detection modules by a couple of lossy channels at each leg of the measurement device. We then work out recursive equations for the end-to-end CM, upon which we perform our finite-key security analysis. 

To continue, we shall first account for the probabilistic (post-selection) nature of the NLAs. Take that in total $N$ signals are transmitted, i.e., assume $N$ runs. The meaning of `run' is well understood in a single-link protocol. It however may be slightly more complex in a repeater setup with essentially  probabilistic links. Here, by each run we mean that TMSV sources at all stations simultaneously transmit a signal. Each signal then has the chance to be successfully amplified by an NLA placed at the other end of the link. In the following, we account for the post-selection effect of the NLAs by referring to~\cite{Pirandola:FS2020}, which has studied a similar post-selection problem in the scope of free-space quantum communications.

Of the overall $N$ runs of the protocol $p_{\rm s}N$, where $p_{\rm s}$ is success probability of the repeater system, will be post-selected by NLAs. In other words, they post-select a portion $p_{\rm s}N$ of the signals. 
Hence, assuming that EC is successful with probability $p_{\rm ec}$, an average number of $np_{\rm s} p_{\rm ec}$ signals contribute to the final key and, therefore, Eq.~\eqref{comprate} takes the form 
\begin{align}
\label{comprate2}
K \geq & \frac{np_{\rm s}p_{\rm ec}}{N} \Big( K_{\rm pe} - \frac{\Delta_{\rm aep}}{\sqrt{np_{\rm s}}} + \frac{\Theta}{np_{\rm s}} \Big).
\end{align}

Before we present numerical results, let us briefly weigh up the action of NLAs. 
Firstly, to see if such devices can be practically useful, we allow some weakening of the Gaussian assumption. This is because the NLA-assisted relays can actually impose non-Gaussianity, which as pointed out earlier, is necessary to (possibly) outdistance the PLOB. As a well-known realization, we can take the action of quantum scissors (QSs), as non-deterministic NLAs, as a guide. Quantum scissors were introduced in \cite{Ralph:QSNLA}, and were studied further in~\cite{Caves:PRA2012,Pandey:QAmps2013}. While the ideal NLA operation is unphysical, in the sense that it works only with zero probability of success, QSs can act as almost-ideal NLAs under weak signal assumptions.
More precisely, it has been  shown that a QS can almost-noiselessly amplify an input coherent state $|\alpha\rangle$ to $|g\alpha\rangle$ with the success probability of a QS $p_{\rm s}=1/g^2$, assuming that $g^2|\alpha|^2\ll 1$~\cite{Pandey:QAmps2013,Xiang:NLA,McMahon:Optimal_QS}. In the prepare-and-measure (P\&M) protocol, where each link has an initial Gaussian modulation variance $V_A$, a similar assumption holds: $\eta g^2 V_A \ll 1$, where we have also take into account the channel loss (We note that the P\&M and entanglement-based protocols are related via $\mu=V_A+1$.).

\begin{figure}[t]
\vspace{+.1cm}
 \includegraphics[scale=0.6]{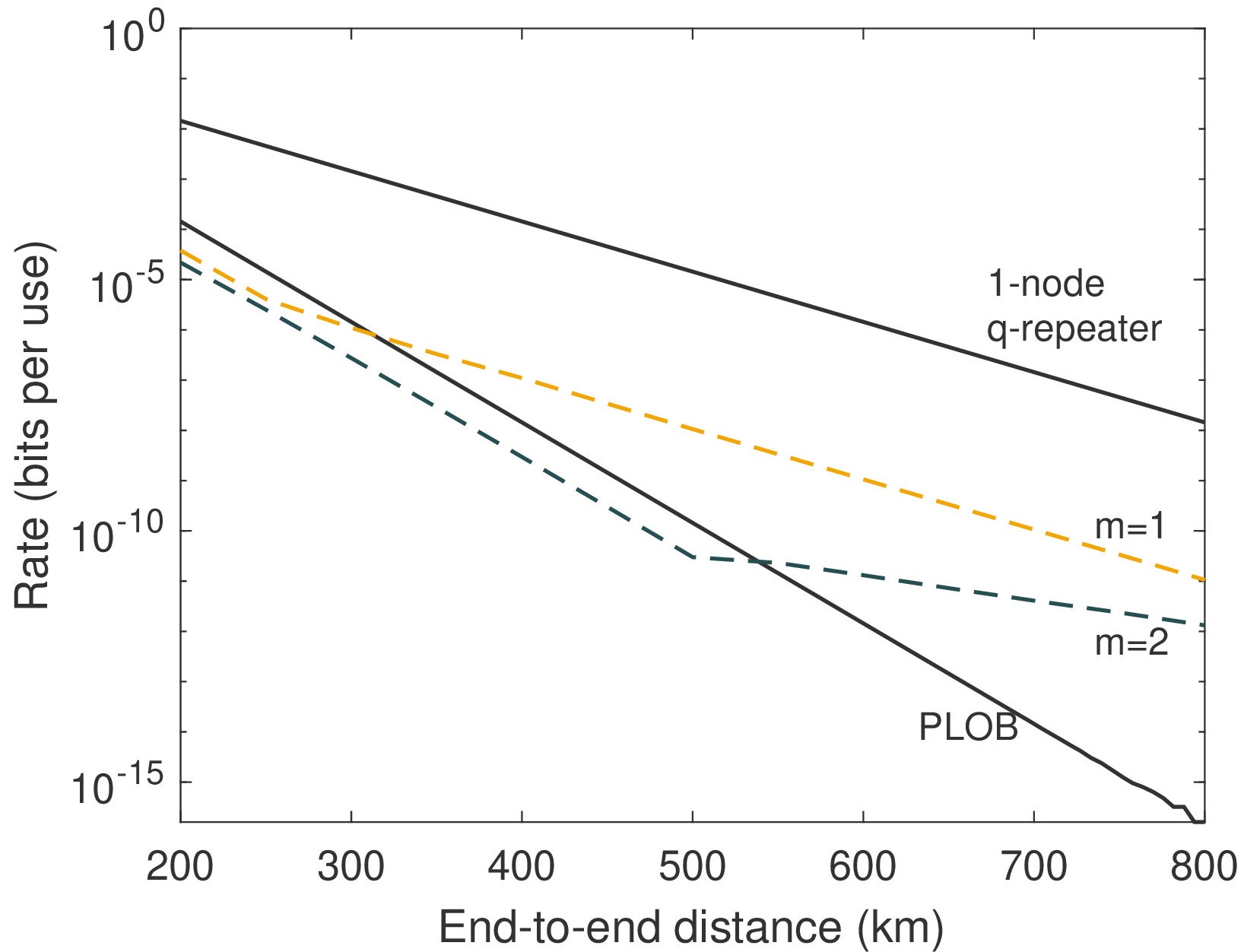}
\caption{{\bf Asymptotic key rate per use of quantum repeater chain.} We consider a heterodyne-based CV-QKD protocol implemented over a quantum repeater chain with depths $m=1$ and $m=2$. Here, we assume reconciliation efficiency $\beta=0.98$, ideal Bell detection modules with $\eta_{\rm B}=1.0$, and zero excess noise $\epsilon=0$ for each single link. Rates are compared with the repeaterless capacity, i.e., PLOB bound~\cite{Pirandola:PLOB2017}, and the single-repeater capacity~\cite{Pirandola:EndtoEnd2019}.}
\label{fig:NLAchain}
\end{figure}

In Fig.~\ref{fig:NLAchain}, by using the recursive equations, we plot the asymptotic secret key rate versus the overall distance between Alice and Bob. Here we have assumed that an ideal chain with $\eta_{\rm B}=1$ and $\epsilon=0$. We encounter a dual optimisation problem, which we solve numerically by optimizing over input modulation $\mu$ and amplification gain $g$, while making sure that $\eta g^2 V_A <10^{-2}$. 

We interpret the results as follows. The curves show that a quantum repeater chain with $m=1$ ($m=2$) can outperform the ultimate benchmarks at about 300 km (500 km), before which the optimized amplification gain is $g=1$, meaning that no amplification in needed. 
Although these results look interesting, when we deviate from the ideal case, i.e., relay loss $\eta<1$ and excess noise $\epsilon>0$ (specifically we could not find a positive rate for, e.g., $\eta_{\rm B}=0.999$ and $\epsilon=0.001$~SNU). 
As discussed for a chain without NLAs, this is partly due to the absolutely symmetric (MDI-like) design that we assumed through our example.  
With a different, possibly asymmetric, design of the repeater links, it may be possible that one can obtain positive rates (nevertheless, the methodology remains the same as presented in this manuscript). From this prospective, our results are the starting point for future studies on NLA-based quantum repeaters.

\bigskip
{\bf\large{Conclusions}}

In summary, we have analysed the composable end-to-end security of Gaussian quantum networks in the presence of generally-untrusted nodes. Assuming two arbitrary end-users of the network, we established a methodology that enables them to complete the crucial task of parameter estimation based only on the data remotely possessed. We have further investigated how they can use the estimated parameters and compute the composable key rate in the finite-size regime. Our study does not need to estimate channel parameters of the individual links that make the route between the two users. In fact, other than being Gaussian, it does not make any assumptions about the communication links, stations, and/or any other components involved. 

Furthermore, {we backed up our theory} by considering the specific case of a chain of {identical} quantum {links, both with and without} NLAs.
In our NLA-assisted design, we assumed ideal NLAs for two reasons. Firstly, {under weak signal assumptions,} they {can be assumed} Gaussian operations~\cite{Ralph:QSNLA,Caves:PRA2012,Pandey:QAmps2013} {(a good example of these NLAs are quantum scissors~\cite{Ralph:QSNLA})}.  
Secondly, since they are ideal, in the sense that they do not add extra noise to the system, they help to obtain the ultimate performance that can be achieved by means of such designs. {While we could show that an NLA-assisted chain can beat the repeaterless limit, we question its practicality. Compared with MDI protocols, we conclude that, apart from noise and loss, this is mostly due to symmetric design of the chain. }  
In addition, for achieving a real-world analysis one can replace the ideal NLAs with realistic ones. This can nevertheless obsolete the Gaussianity of the network so this next step will have to be investigated cautiously.

\bigskip
{\bf\large{Methods}}

{\bf The worst-case scenario covariance matrix.}
In the following, we discuss the confidence intervals for $\widehat{\boldsymbol{\Sigma}}$. Despite the fact that this procedure is based on the shared data, it can have a direct application on a theoretical CM as in Eq.~(\ref{CM:recursive}) defined through a specific model of the links between the parties.
Our analysis relies on tail bounds for the chi-squared distribution. Assume that random data variables $[q]_1,\dots, [q]_N$ from the variable $q$ which follows a normal distribution with unit variance and zero mean value. 
Then, the random variable 
\begin{align}
Q=\sum_{i=1}^{m_{\rm pe}} [q]^2_i \sim \chi^2(m_{\rm pe},0)
\end{align}
follows a chi-squared distribution and allows for the following tail bounds~\cite[Lemma~6]{tailbounds}
\begin{align}
P[Q\geq m_{\rm pe}+2\sqrt{m_{\rm pe} \kappa}+2 \kappa]&\leq \exp(-\kappa), \label{eq:tailbounds1}\\
P[Q\leq m_{\rm pe} -2\sqrt{m_{\rm pe} \kappa}]&\leq \exp(-\kappa), \label{eq:tailbounds2}
\end{align}
where $\kappa$ is related to the error of PE, $\varepsilon_{\rm pe}$, as we shall see shortly. 

From the samples $[q_x]_i$ and $[q_y]_i$, one can define standard normal variables $[\mathsf{q}_x]_i=[q_x]_i/\sqrt{\sigma_x^2}$ and $[\mathsf{p}_\mathsf{y}]_i=[p_y]_i/\sqrt{\sigma_y^2}$ so the estimators of $\langle q_x^2 \rangle$ and $\langle q_y^2 \rangle$ can be expressed as
\begin{align}\label{eq:est_var}
\widehat{\langle q_x^2\rangle}:=m_{\rm pe}^{-1}\sum_{i=1}^{m_{\rm pe}} [q_x]_i^2= & \sigma_x^2 m_{\rm pe}^{-1}\sum_{i=1}^{m_{\rm pe}}[\mathsf{q}_x]^2_i , \\
\widehat{\langle q_y^2\rangle}:=m_{\rm pe}^{-1}\sum_{i=1}^{m_{\rm pe}} [q_y]_i^2= & \sigma_x^2 m_{\rm pe}^{-1}\sum_{i=1}^{m_{\rm pe}}[\mathsf{q}_y]^2_i,
\end{align}
where $\sum_{i=1}^{m_{\rm pe}}[\mathsf{q}_x]^2_i$ and $\sum_{i=1}^{m_{\rm pe}}[\mathsf{q}_y]^2_i$ are chi-square variables following the tail bounds: 
%\begin{align}
%\sum_{i=1}^{m_{\rm pe}}[\mathsf{q}_x]^2_i>{m_{\rm pe}}+2\sqrt{{m_{\rm pe}}\kappa}+2\kappa, \\ \sum_{i=1}^{m_{\rm pe}}[\mathsf{q}_y]^2_i>{m_{\rm pe}}+2\sqrt{{m_{\rm pe}}\kappa}+2\kappa. 
%\end{align} 
\begin{equation}
\min \left[  \sum_{i=1}^{m_{\rm pe}}[\mathsf{q}_x]^2_i , \sum_{i=1}^{m_{\rm pe}}[\mathsf{q}_y]^2_i  \right] >{m_{\rm pe}}+2\sqrt{{m_{\rm pe}}\kappa}+2\kappa.
\end{equation}
This guarantees that maximum noise is considered based on the data, and implies that 
\begin{align}\label{eq:wcs_var}
  \widehat{\langle q_x^2\rangle} \geq\langle q_x^2 \rangle_{\rm wc}, ~  \widehat{\langle q_y^2\rangle} \geq\langle q_y^2 \rangle_{\rm wc}, 
\end{align}
with probability
\begin{align}
   \text{Pr}\left[\widehat{\langle q_x^2\rangle} \geq\langle q_x^2 \rangle_{\rm wc}\right] & \leq \exp(-\kappa), \\
  \text{Pr}\left[ \widehat{\langle q_y^2\rangle}\geq\langle q_y^2 \rangle_{\rm wc}\right] & \leq \exp(-\kappa),
\end{align}
for the worst-case scenario values
\begin{align}
\langle q_x^2 \rangle_{\rm wc}= & \sigma_x^2{m_{\rm pe}}^{-1}({m_{\rm pe}}+2\sqrt{{m_{\rm pe}}\kappa})+\mathcal{O}(1/{m_{\rm pe}}) \notag \\ = & \sigma_x^2(1+2\sqrt{\kappa/{m_{\rm pe}}}),\label{eq:variance1}\\
\langle q_y^2 \rangle_{\rm wc}= & \sigma_y^2{m_{\rm pe}}^{-1}({m_{\rm pe}}+2\sqrt{{m_{\rm pe}}\kappa})+\mathcal{O}(1/{m_{\rm pe}}) \notag \\ 
= & \sigma_y^2(1+2\sqrt{\kappa/{m_{\rm pe}}}).\label{eq:variance2}
\end{align}

To find the worst-case scenario values for the covariance term $\langle xy\rangle$ we make the following calculations: Combining the samples $[x]_i$ and $[y]_i$ accordingly, we have that
\begin{align}
([q_y]_i-[q_x]_i)^2=&[q_y]_i^2+[q_x]_i^2-2[q_y]_i[q_x]_i,\\
([q_y]_i+[q_x]_i)^2=&[q_y]_i^2+[q_x]_i^2+2[q_y]_i[q_x]_i,
\end{align}
which leads to the relation
\begin{align}
[q_y]_i[q_x]_i=\frac{1}{4}\left[([q_y]_i+[q_x]_i)^2-([q_y]_i-[q_x]_i)^2\right].
\end{align}
The variables $[q_y]_i\pm [q_x]_i$ are zero-mean Gaussian variables with variances $V_\pm$ since $[q_x]_i$ and $[q_y]_i$ are assumed to be Gaussian. 
More specifically, the variables $[q_{z_\pm}]_i=([q_y]_i\pm [q_x]_i)/\sqrt{V_\pm}$ are standard normal variables. Thus by summing over all the samples and dividing by ${m_{\rm pe}}$, we may express the estimator of $\langle q_xq_y \rangle$ as 
\begin{align}
    & \widehat{\langle q_xq_y\rangle}  : ={m_{\rm pe}}^{-1}\sum_{i=1}^{m_{\rm pe}} [q_y]_i[q_x]_i \notag \\
    & =\frac{1}{4}\left(V_+{m_{\rm pe}}^{-1}\sum_{i=1}^{m_{\rm pe}}[q_{z_+}]^2-V_- {m_{\rm pe}}^{-1}\sum_{i=1}^{m_{\rm pe}}[q_{z_-}]^2\right),
\end{align}
where $\sum_{i=1}^{m_{\rm pe}}[q_{z_\pm}]^2$ are chi-square distributions following the tail bounds of Eqs.~(\ref{eq:tailbounds1}) and~(\ref{eq:tailbounds2}).

We then impose that the estimator is smaller than its worst-case scenario value $\langle q_xq_y\rangle_{\rm wc}$, i.e.,
\begin{align}\label{eq:wcs_covar}
 \widehat{\langle q_xq_y\rangle} <\langle q_xq_y \rangle_{\rm wc}, % :=   \frac{1}{{m_{\rm pe}}}\sum_{i=1}^{m_{\rm pe}} [q_y]_i[q_x]_i<\langle q_xq_y \rangle_{\rm wc},
\end{align}
where $\langle q_xq_y \rangle_{\rm wc}$ is computed by replacing $\sum_{i=1}^{m_{\rm pe}}[q_{z_\pm}]^2$ with the tail bounds in Eqs.~(\ref{eq:tailbounds1}) and~(\ref{eq:tailbounds2}),
i.e., using 
%%%%%%%%%%%%%%%%%%%%%%%%%%%%%%%%%%%%%%%%%%%%%%%%%%%%%%%%%%% which guarantees the least amount of correlation based on the data, and implies that at least one of the following terms
\begin{align}
    \sum_{i=1}^{m_{\rm pe}}[q_{z_+}]^2<{m_{\rm pe}}-2\sqrt{{m_{\rm pe}} \kappa}, \label{SP1}
\end{align}
and
\begin{align}
\sum_{i=1}^{m_{\rm pe}}[q_{z_-}]^2>{m_{\rm pe}}+2\sqrt{{m_{\rm pe}} \kappa}+2\kappa. \label{SP2}
\end{align}
Therefore, up to $\mathcal{O}(1/{m_{\rm pe}})$,  we have 
\begin{align}\label{eq:covariance}
\langle q_xq_y\rangle_{\rm wc}= & \frac{1}{4}\Bigg(V_+ \frac{1}{{m_{\rm pe}}}({m_{\rm pe}}-2\sqrt{{{m_{\rm pe}}}\kappa}) \notag\\&- V_-\frac{1}{{m_{\rm pe}}}({m_{\rm pe}}+2\sqrt{{m_{\rm pe}}\kappa}+2\kappa)\Bigg) \notag\\
%= & \frac{1}{4}\Bigg(V_+(1-2\sqrt{\kappa/{m_{\rm pe}}})\notag\\&
% - V_-(1+2\sqrt{\kappa/{m_{\rm pe}}})\Bigg) \notag\\
= & \frac{1}{4}\left((V_+-V_-)-2\sqrt{\kappa/{m_{\rm pe}}}(V_++V_-)\right)\notag\\
= & \langle q_x q_y\rangle -\sqrt{k/{m_{\rm pe}}}(\langle q_x^2\rangle+\langle q_y^2 \rangle).
\end{align}

Note that a necessary condition for $\widehat{\langle q_x p_y\rangle}<\langle q_xq_y\rangle_{\rm wc}$ to be valid is that either Eq.~(\ref{SP1}) or Eq.~(\ref{SP2}) is valid. Therefore
\begin{align}\label{eq:prob_cor}
&\text{Pr}\left[ \widehat{\langle q_x p_y\rangle}<\langle q_xq_y\rangle_{\rm wc}\right]\notag\\& \le \text{Pr}\left[\left(\sum_{i=1}^{m_{\rm pe}}[q_{z_+}]^2<{m_{\rm pe}}-2\sqrt{{m_{\rm pe}} \kappa}\right)\right.\notag\\
&\left.\vee \left(\sum_{i=1}^{m_{\rm pe}}[q_{z_-}]^2>{m_{\rm pe}}+2\sqrt{{m_{\rm pe}} \kappa}+2\kappa\right)\right] \notag\\
 &\leq \text{Pr}\left[\left(\sum_{i=1}^{{m_{\rm pe}}_{\rm pe}}[q_{z_+}]^2<{m_{\rm pe}}-2\sqrt{{m_{\rm pe}} \kappa}\right)\right]\notag\\&+\text{Pr}\left[\left(\sum_{i=1}^{m_{\rm pe}}[q_{z_-}]^2>{m_{\rm pe}}+2\sqrt{{m_{\rm pe}} \kappa}+2\kappa\right)\right] \notag\\
    &\leq 2\exp(-\kappa) .
\end{align}

Similarly, the parties calculate equivalent relations for the data from the $p$-quadrature. They obtain corresponding equations for $\langle p_x^2 \rangle_{\rm wc}$, $\langle p_y^2 \rangle_{\rm wc}$, and $\langle p_xp_y \rangle_{\rm wc}$ following Eqs.~(\ref{eq:variance1}),~(\ref{eq:variance2}), and~(\ref{eq:covariance}). In particular, since $\langle p_xp_y \rangle_{\rm wc}$ is a negative quantity, the corresponding probability  of Eq.~\eqref{eq:prob_cor} will have as an argument an inequality with a different direction and the minus sign in the corresponding Eq.~\eqref{eq:covariance} will be replaced by a plus sign. 

All the worst-case parameters $\langle ... \rangle_{\mathrm{wc}}$ define the worst-case scenario CM $\boldsymbol{\Sigma}_{\rm wc}$ which has the form of Eq.~(\ref{SP3}) of the main text but with
the replacements $\langle ... \rangle  \rightarrow \langle ... \rangle_{\mathrm{wc}}$. From the previous derivations, we see that
\begin{align} 
\boldsymbol{\Sigma}_{\rm wc}=\widehat{\boldsymbol{\Sigma}}+\sqrt{\frac{4\kappa}{{m_{\rm pe}}}}
\begin{pmatrix}
\widehat{\mathbf{V}}_x & -\frac{\widehat{\mathbf{V}}_x+\widehat{\mathbf{V}}_y}{2}\mathbf{Z}\\
-\frac{\widehat{\mathbf{V}}_x+\widehat{\mathbf{V}}_y}{2}\mathbf{Z}&\widehat{\mathbf{V}}_y
\end{pmatrix},
\end{align}
where $\widehat{\boldsymbol{\Sigma}}$ is exactly the one defined in Eq.~(\ref{SP3}) of the main text, together with the associated $\widehat{\mathbf{V}}_x$ and $\widehat{\mathbf{V}}_y$. As we see, the diagonal (noise) terms are increased whereas the off-diagonal (correlation) terms are decreased in modulus. This vanishes in the asymptotic case where ${m_{\rm pe}} \rightarrow \infty$. 

Now, let us assume that at least one of the inequalities in Eqs.~(\ref{eq:wcs_var}) or (\ref{eq:wcs_covar}) is true which happens with total probability $\leq 4\exp(-\kappa)$. Considering
the $p$ quadrature, the total probability of failure is $\leq 8\exp(-\kappa)$. The latter is therefore a bound on the probability that the CM is worse than the worst-case expression 
$\boldsymbol{\Sigma}_{\rm wc}$ (in which case the rate would be less than the worst-case value).  
%\begin{widetext}
%\begin{align}\label{eq:prob_bound}
%&\text{Pr}\left[\left(\frac{\widehat{\langle q_x p_y \rangle}}{\langle q_xq_y\rangle_{\rm wc}}<1\right)\vee  \left(\frac{\widehat{\langle q_x^2\rangle}}{\langle q_x^2 \rangle_{\rm wc}} %>1\right)\vee\left(\frac{\widehat{\langle q_y^2\rangle}}{\langle q_y^2 \rangle_{\rm wc}}>1\right)\right]\notag\\& \leq\text{Pr}\left[\frac{\widehat{\langle q_x p_y \rangle}}{\langle q_xq_y\rangle_{\rm wc}}<1\right]+  %\text{Pr}\left[\frac{\widehat{\langle q_x^2\rangle}}{\langle q_x^2 \rangle_{\rm wc}} >1\right]+  \text{Pr}\left[\frac{\widehat{\langle q_y^2\rangle}}{\langle q_y^2 \rangle_{\rm wc}} >1\right]\notag\\&
%\leq 4\exp(-\kappa).
%\end{align}
%\end{widetext}
%Consider now that we have the same relation for the data associated with the $p$-quadrature.
%Then $\boldsymbol{\Sigma}_{\rm wc}$ is not the worst-case CM according to this data; hence, the PE step will be assumed failed. 
The parties can only allow this to happen with a very small probability that is less than $\varepsilon_{\rm pe}$. Therefore, by bounding the previous relation we have that
\begin{align}
8\exp(-\kappa) & \leq \varepsilon_{\rm pe}, 
\end{align}
which defines $\kappa=\text{ln}(8/\varepsilon_{\rm pe})$. %Note that here we have considered the double of the bound $4 \exp(-\kappa)$ of Eq.~(\ref{eq:prob_bound}) taking into account both quadratures.

Finally, for calculating the secret key rate of Eq.~\eqref{comprate2} from the theoretical CM $\mathbf{V}$ in Eq.~\eqref{CM:recursive} we apply the inverse transformation of Eq.~\eqref{CM:data}.
%for $\mathbf{L}$ being the identity matrix since Alice is preparing the the signal states by applying a measurement on an entangled mode to the traveling mode. 
In this way, we obtain a theoretical version of the classical CM
\begin{align}
\boldsymbol{\Sigma}^\text{thr}& =   \left[\mathbf{V}+(\mathbf{I}\oplus\mathbf{I})\right]
:=\begin{pmatrix}
\mathbf{V}^\text{thr}_x&\mathbf{C}^\text{thr}_{xy}\\
\mathbf{C}^\text{thr}_{xy}&\mathbf{V}^\text{thr}_y
\end{pmatrix}.
\end{align}
By using this CM, we can calculate the worst-case scenario theoretical CM
\begin{align}
\boldsymbol{\Sigma}^\text{thr}_{\rm wc}=\boldsymbol{\Sigma}^\text{thr}+\sqrt{\frac{4\kappa}{m_{\rm pe}}}
 \begin{pmatrix}
\mathbf{V}^\text{thr}_x & -\frac{\mathbf{V}^\text{thr}_x+\mathbf{V}^\text{thr}_y}{2}\mathbf{Z}\\
-\frac{\mathbf{V}^\text{thr}_x+\mathbf{V}^\text{thr}_y}{2}\mathbf{Z}&\mathbf{V}^\text{thr}_y
\end{pmatrix}. 
\end{align}
Then, we calculate the worst-case scenario theoretical quantum CM using
\begin{align}
    \mathbf{V}^\text{thr}_\text{wc}= \boldsymbol{\Sigma}^\text{thr}_{\rm wc}-(\mathbf{I}\oplus\mathbf{I}).
\end{align}
The latter is finally used to compute $K_\text{pe}$ of the composable secret key rate in Eq.~\eqref{comprate2} by following the steps~\eqref{cond_cm}-\eqref{mutual_info}. %\PP{Note that here $\mathbf{L}$ is given according to the relation below Eq.~(\ref{CM:data}) with $\mu$ being equal to $\mu=a_0$ in Eq.~(\ref{CM:recursive}).}

\smallskip
{\bf\large{Acknowledgements}}

This work has been funded by the European Union via \textquotedblleft Continuous Variable Quantum Communications\textquotedblright\ (CiViQ, Grant Agreement No. 820466)
and the EPSRC via the UK Quantum Communications Hub (Grant No. EP/T001011/1).

\bibliography{references}

\end{document}